\documentclass[%
 aip,
 jap,%
 amsmath,amssymb,
preprint,%
]{revtex4-1}

\usepackage{graphicx}
\usepackage{dcolumn}
\usepackage{bm}
\usepackage{color}
\usepackage{multirow}
\usepackage{balance}
\makeatletter

\newcommand{\Rmnum}[1]{\expandafter\@slowromancap\romannumeral #1@}
\makeatother

\begin{document}

\preprint{AIP/123-QED}

\title{Exchange Bias and Memory Effect in Double Perovskite Sr$_2$FeCoO$_6$}

\author{Pradheesh R.*}
\affiliation{Low Temperature Physics Laboratory, Department of Physics, Indian Institute of Technology Madras, Chennai 600036, India.}
\author{Harikrishnan S. Nair*}
\affiliation{J\"{u}lich Center for Neutron Sciences-2/Peter Gr\"{u}nberg Institute-4, Forschungszentrum J\"{u}lich, 52425 J\"{u}lich, Germany}
\author{V. Sankaranarayanan}
\affiliation{Low Temperature Physics Laboratory, Department of Physics, Indian Institute of Technology Madras, Chennai 600036, India.}
\author{K. Sethupathi}
\affiliation{Low Temperature Physics Laboratory, Department of Physics, Indian Institute of Technology Madras, Chennai 600036, India.}

\date{\today}
\begin{abstract}
We report on the observation of exchange bias and {\it memory effect} in
double perovskite Sr$_2$FeCoO$_6$.
Antiphase boundaries between the ferromagnetic
and antiferromagnetic regions in the disordered glassy phase
is assumed as responsible for the observed effect
which reflects in the cooling field dependence and temperature evolution of
exchange bias field and in {\it training effect}.
The spin glass phase itself is characterized through {\it memory},
{\it ageing} and magnetic relaxation experiments.
The spin glass transition temperature, $T_g$, versus $H_{dc}^{2/3}$
follows the Almeida-Thouless line yielding a freezing temperature,
$T_f$ = 73~K.
Time-dependent magnetic relaxation studies reveal the magnetization dynamics
of the underlying glassy phase in this double perovskite.
\end{abstract}
\maketitle
\begin{figure}
\centering
\includegraphics[scale=0.65]{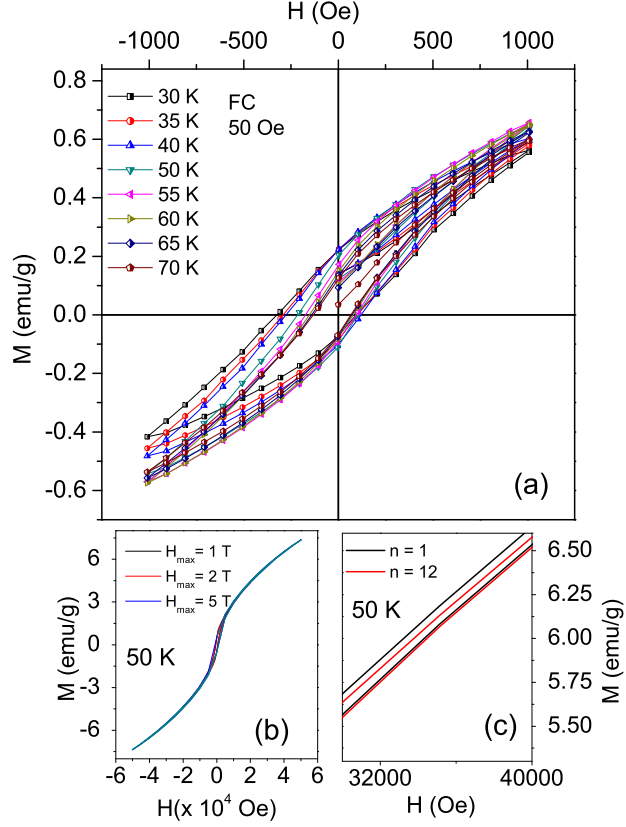}
\caption{(colour online) (a) Field-cooled (50~Oe) isothermal magnetization of Sr$_2$FeCoO$_6$ at different temperatures below $T_g$. The field range is limited to $\pm$1000~Oe ($H_{max}$) which is greater than $H_A$ (i.e., $H_{max} > H_A$). (b) Hysteresis plots at different $H_{max}$ where exchange bias disappears. (c) {\it Training effect} at 50~K observed with 12 loops.}
\label{MHT}
\end{figure}
\begin{figure}
\centering
\includegraphics[scale=0.60]{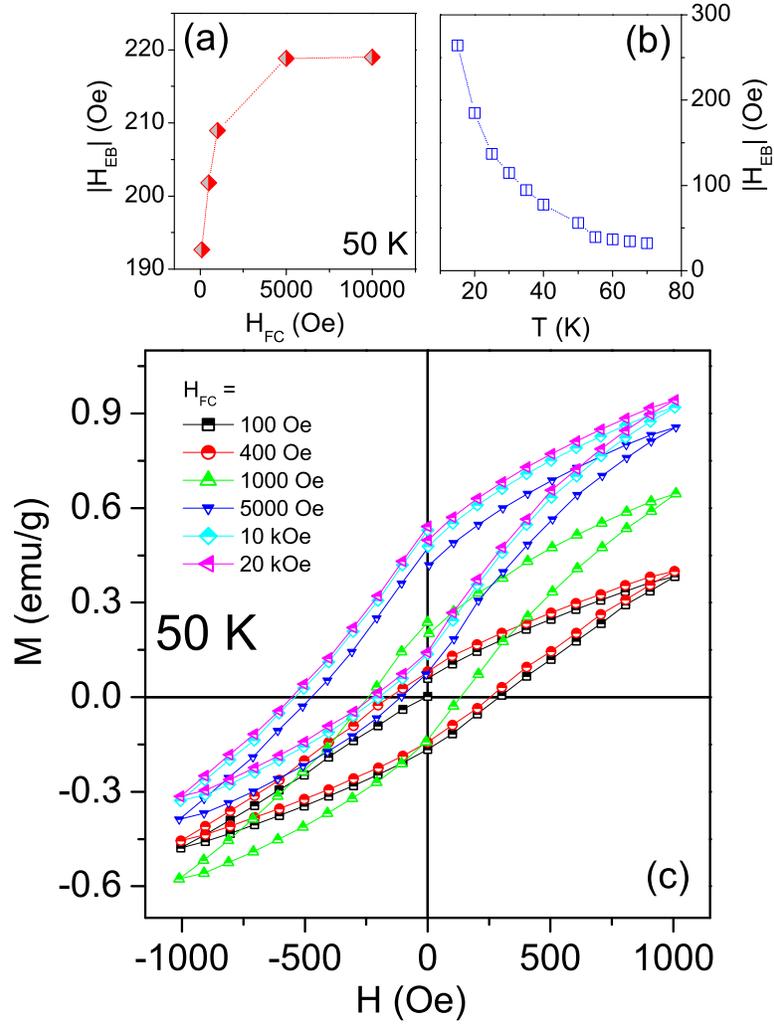}
\caption{(colour online) Exchange bias field $H_{EB}$ plotted against (a) $H_{FC}$ and (b) temperature conform to typical exchange bias characteristics. (c) The field-cooled hysteresis curves at different cooling fields up to 20~kOe display vertical displacement that signify FM clusters present in the system.}
\label{MH}
\end{figure}
\begin{figure}
\centering
\includegraphics[scale=0.65]{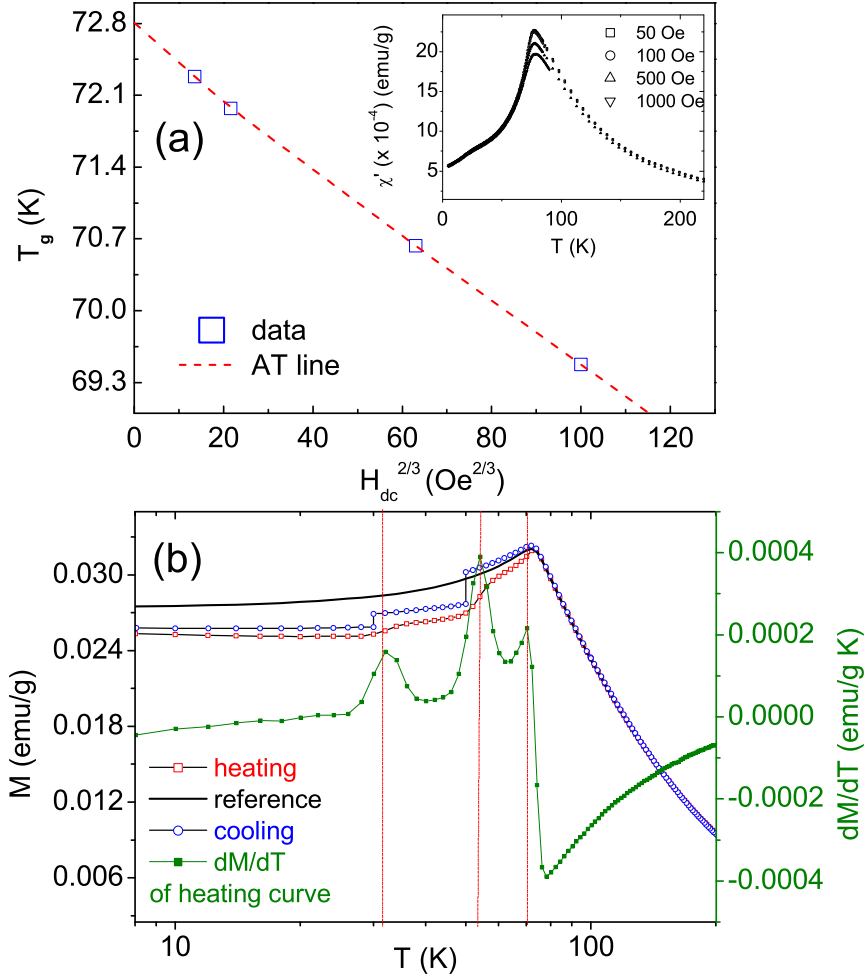}
\caption{(colour online) (a) $T_g$ vs $H^{2/3}_{dc}$ plot for Sr$_2$FeCoO$_6$
where $H_{dc}$ is the superimposed dc field and the dashed-line shows the
fit according to Eqn.~\ref{AT}. The inset shows the real part of ac susceptibility, $\chi'(T)$,
at different applied dc fields. (b) {\it Memory} effect in Sr$_2$FeCoO$_6$. The stops administered during the cooling curve, where the measurement is halted, are recovered in the heating cycle. The derivative plots clearly show the recovered stops. For comparison, a reference measurement curve (without stops) is also presented.}
\label{AT_memory}
\end{figure}
\begin{figure}
\centering
\includegraphics[scale=0.65]{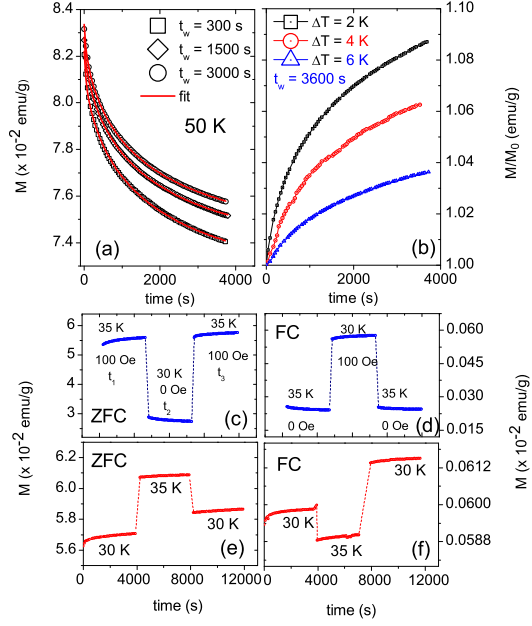}
\caption{(colour online) (a) Time-dependent magnetization of Sr$_2$FeCoO$_6$ at 50~K for three different wait times. The solid lines are fit to the eqn~(\ref{K}). (b) The time-dependent magnetization with a wait time of $t_w$ = 3600~s. Signatures of {\it aging} are present in relaxation experiments in both ZFC ((c), (e)) and FC ((d), (f)) protocols. (c) and (d) show negative temperature cycling in ZFC and FC mode respectively while (e) and (f) show the same plots in warming mode. Asymmetric response is clearly seen for the warming mode.}
\label{relax}
\end{figure}

%
%
Materials that show vertical/horizontal displacement of
magnetic hysteresis loop -- known as exchange bias materials --
are potential candidates for
technological applications as spin valves
\cite{park_2011},
permanent magnets
\cite{nogues_1999}
and in magnetic recording
\cite{liao_2010, roshchin_2010}.
Exchange bias is observed mainly in
ferromagnetic (FM)/ antiferromagnetic (AFM) bilayers
\cite{stiles_2001}
but, also in nanoparticles
\cite{skumryev_2003},
inhomogeneous magnets,
\cite{nogues_1999}
and strongly correlated oxides
like manganites,
\cite{karmakar_prb_77_144409_2008}
cobaltites,
\cite{tang_prb_73_174419_2006}
and in intermetallics
\cite{chen_prb_72_054436_2005}.
In classical exchange bias (EB) systems, the hysteresis loop is
shifted to the left of the origin and conventionally EB is defined negative.
Positive EB has also been reported, for example, in metallic bilayers
\cite{nogues_prl_76_4624_1996}
and in spin glasses
\cite{ali_nature_6_70_2006}.
In a detailed study to distinguish between reentrant spin glass (RSG)
and cluster glass (CG), exchange bias with a shift in both magnetization
and field axis was observed in L$_{0.5}$Sr$_{0.5}$CoO$_3$
\cite{mukherjee_1996}.
In this Letter, we report the observation of exchange bias in a spin glass
double perovskite thereby, extending the generality of this phenomenon.
\\
%
The spin glass (SG) nature of Sr$_2$FeCoO$_6$, with transition temperature $T_g \approx$ 75~K,
studied through macroscopic magnetization and structural studies
using neutrons has been reported elsewhere
\cite{pradheesh}.
In this paper we focus on detailed magnetization measurements in field-cooled
and zero field-cooled conditions along with magnetic relaxation measurements
conducted using commercial SQUID magnetometer
and physical property measurement system (both M/s Quantum Design Inc.).
\\
As the first set of magnetization measurements, field-cooled hysteresis curves at different
temperatures were measured on Sr$_2$FeCoO$_6$.
To this effect, the sample was field-cooled from 120~K to a
temperature below $T_g$ with an applied field of 50~Oe
(after each $M (H)$ curve the sample was demagnetized by warming up to 120~K).
The field-cooled magnetic hysteresis loops at different temperatures in the range
30 -- 70~K, that show clear shifts resembling EB
are presented in Fig~\ref{MHT} (a).
The loop-shifts in the $M (H)$ plots, as seen in the figure,
can signify exchange bias due to the spins at
of FM/AFM, FM/SG interfaces.
In Sr$_2$FeCoO$_6$, antisite disorder leads to SG phase at low temperature
which then forms FM/SG interfaces which can cause exchange anisotropy.
In order to avoid minor loop effect in the observation of a genuine EB-shift, the optimal maximum applied field ($H_{max}$) should be greater than the anisotropy field (H$_A$) of the system.
From the analysis of initial magnetization at 50~K using
$M$ = $M_s(1 - a/H - b/H^2) + \chi H$ and using the relations
$b$ = $4K^2_1/15M^2_s$ and $H_A$ = $2K_1/M_s$
\cite{andreev_jac_260_196_1997,patra_jpcm_21_078002_2009},
a rough estimate of the anisotropy field
$H_A$ = 448~Oe was obtained ($M_s$ is saturation magnetization,
$a$, $b$ are free-parameters, $K_1$ is anisotropy constant, $\chi$
is the high-field susceptibility).
Consequently, our hysteresis measurements were performed
such that the maximum applied field $H_{max} > H_A$.
Fig.~\ref{MHT} (b), shows that the effect of applied
fields greater than 10~kOe is to diminish the effect of exchange bias.
Similar effect of vanishing EB at high fields has been reported for
cluster glass perovskite cobaltites
\cite{luo_2007}.
In order to confirm that the exchange bias effect is intrinsic, we performed
{\it training effect} experiment where the $M (H)$ curve at 50~K is recorded in field-cooled condition for 12 continuous loops.
In Fig~\ref{MHT} (c), a magnified view of the 1$^{st}$ and 12$^{th}$ loops are presented
showing a clear shift which is typical of the response from exchange biased systems
and hints at the metastable nature of the interface
\cite{giri_jpcm_23_073201_2011}.
Field-cooled hysteresis curves at 50~K were measured as a function of
different cooling fields, $H_{FC}$.
The exchange bias field was estimated from the
$M (H)$ loops as, $|H_{EB}|$ = $|(H_+ + H_-)/2|$;
$H_+$ and $H_-$ are the positive and negative intercepts of the magnetization
curve with the field axis.
The variation of $H_{EB}$ as a function of $H_{FC}$
and temperature are shown in Fig.~\ref{MH} (a) and (b) respectively.
In order to probe RSG features in the present system, we performed field-cooled hysteresis measurements at different cooling fields as suggested in Mukherjee et. al.\cite{mukherjee_1996}.
The resulting $M (H)$ plots are presented in Fig~\ref{MH} (c).
Displaced hysteresis loops are evident which, with increasing
$H_{FC}$ intersect the $H$-axis at a progressively higher negative values.
Similar feature has been observed in La$_{0.5}$Sr$_{0.5}$CoO$_3$
\cite{mukherjee_1996}
and signifies the presence of FM clusters.
\\
In the case of spin glass systems, the peak at $T_g$ in the imaginary part of ac
magnetic susceptibility shifts to low temperature with increasing value of
superimposed dc field.
The evolution of $T_g$ with applied magnetic field can be
explained by Almeida-Thouless ($AT$) line in a $H$-$T$ phase diagram
\cite{almeida_1987}.
The $AT$ line is described by the equation,
\begin{equation}
\frac{H_{dc} (T)}{\Delta J} = (1-T_g/T_C)^{\alpha}
\label{AT}
\end{equation}
Here $\Delta J$ is the width of the distribution of exchange interactions,
$H_{dc}$ is the superimposed dc magnetic field and T$_C$ is the transition temperature.
Fig~\ref{AT_memory} (a) is the plot of $T_g$ versus $H^{2/3}_{dc}$ for Sr$_2$FeCoO$_6$
which shows a decrease in $T_g$ with applied field.
A fit to the variation of $T_g$ with field assuming the
critical exponent $\alpha= 3/2$ gives straight line fit satisfying
the $AT$ equation and confirms the spin glass nature of Sr$_2$FeCoO$_6$.
Linear behaviour with the critical exponent being $3/2$ is observed for low fields ($\mu_0$H $<$ 1~T).
Extrapolating the fit to both the axes, we obtain
the freezing temperature ($T_f$) and the critical
field ($H_{cr}$) as 73~K and 116~Oe respectively.
The conformity with $AT$-line has been observed in
intrinsically exchange biased Zn$_x$Mn$_{3-x}$O$_4$
solid solutions
\cite{shoemaker_prb_80_144422_2009}.
For reentrant spin glass systems, the disorder extends to the
whole volume resulting in the shift of $T_f$ to lower
temperatures with increase in magnetic field
\cite{martinez_1998}.
\\
In order to further test the glassy magnetic ground state
in Sr$_2$FeCoO$_6$, {\it memory} experiments were conducted,
for which, the sample is first zero field-cooled
to low temperatures at a constant cooling rate,
while recording magnetization.
While cooling, intermediate stops are administered
below $T_c$ when the measurement is stopped (for 2~h)
and magnetization is allowed to relax.
After reaching the lowest possible temperature,
the sample is heated back at a constant heating rate
without administering any stops and magnetization is
recorded.
For comparison, a reference curve where no stops are administered is also
recorded.
The cooling, heating, reference
and the derivative of the heating curve, $dM/dT$, are
presented in Fig~\ref{AT_memory} (b).
Clear {\lq \it dips\rq} in magnetization of the cooling curve are discernible
where the measurement was stopped.
The signatures of {\it memory} effect is clear in the heating
cycle where the steps are recovered at the same temperature points.
This is clearly visible in the plot of $dM/dT$.
Observation of {\it memory} effect is a confirmation of the
magnetic glassy state and has been reported in canonical spin glasses and
phase separated manganites that show spin glass-like ground states.
\\
In order to study the magnetic relaxation mechanisms
stemming from the underlying magnetic glassy state,
time dependent magnetization with different wait times
were recorded at 50~K.
For these measurements, the sample was zero field-cooled
to 50~K, a wait time $t_w$ was administered and then the
magnetization was measured as a function of time.
Evident from Fig~\ref{relax} (a), a clear dependence on $t_w$
can be seen wherein the system
becomes magnetically stiffer as the wait time increases;
which is common among canonical spin glasses.
Fig~\ref{relax} (b) shows time-dependent magnetization
at $t_w$ = 3600~s but at different temperatures.
The time-dependent magnetization was fitted well
with a stretched exponential of the form
\begin{equation}
M(t) = M_0 - M_g exp [-(\frac{t}{t_r})^{1-n}]
\label{K}
\end{equation}
where $M_0$ is the intrinsic ferromagnetic moment, $M_g$ is the
glassy component of the moment, $t_r$ is the characteristic time
component and $n$ is the stretched exponential exponent.
Eqn.~(\ref{K}) is similar to Kohlrausch law
\cite{palmer_1984}
which is used to explain magnetic,
dielectric and optical phenomena where relaxation mechanisms play a important
role in the dynamics.
The exponent in Eqn.~(\ref{K}) is temperature dependent,
and according to the percolation model, varies in the range $\frac{1}{3}\leq~n~\leq 1$
\cite{bohmer_1993}.
Table~\ref{fit1} shows the parameters obtained from the fit
according to Eqn~(\ref{K}).
The values of $n$ and $M_0$ are independent of $t_w$
while the characteristic time scale $t_r$ varies with $t_w$
typical of canonical spin glasses
\cite{chamberlin_1984}.
The characteristic time varies slowly with the wait time, but the dependence of $t_r$ on $t_w$ implies that it is in a non-equilibrium state
and that of the {\it memory effect}.
Negative temperature cycling of magnetization in zero field-cooled (ZFC) and
field-cooled (FC) protocols were also performed to complement the {\it memory effect}.
In the ZFC protocol, the sample was cooled down to 35~K in ZFC
mode and a field of 100~Oe was applied to measure magnetization for time $t_1$.
Further, the sample was cooled down to 30~K, the field is switched off, immediately after which the magnetization is recorded for another time $t_2$.
After $t_2$ the system is taken back to 35~K and magnetization
measured for time $t_3$.
In the above measurement, $t_1$ = $t_2$ = $t_3$ = 3600~s.
Fig~\ref{relax} (c) shows the ZFC temperature cycling where
the effect of {\it memory} is observable even after {\it aging} at lower
temperature.
Fig~\ref{relax} (d) illustrates the same experiment as in (c) but, in FC protocol.
Similar measurements (ZFC and FC) but in heating cycle were performed the results of which are given in Fig~\ref{relax} (e) and (f).
An asymmetric response is observed which means that there is no {\it memory} while heating the sample.
\\
The relaxation measurements confirms the spin glass nature of Sr$_2$FeCoO$_6$ and can be concluded that the observed exchange bias is seen in the spin glass phase.
Exchange bias systems with FM/SG interfaces are known to show exponential decrease in $H_{EB}$ with temperature
\cite{karmakar_prb_77_144409_2008}.
In such a scenario, the SG phase forms the frozen phase
where magnetization is irreversible while that of the FM phase is reversible.
In the present case, we assume a minority FM phase which coexist in a majority
AFM disordered (glassy) phase as deduced from the hysteresis curves and lack of saturation magnetization.
With reduction in temperature, increasing number of disordered AFM (or glassy)
domains freeze and this progressive freezing leads to enhancement of
$H_{EB}$ at low temperature.
The dependence of $H_{EB}$ on $H_{FC}$ is understood based on the competition
between the Zeeman coupling energy and the exchange energies at the interfaces.
At low cooling fields, field-cooling induces progressive enhancement of freezing
of domains and hence $H_{EB}$ increases whereas at high cooling fields,
Zeeman energy overcomes the magnetic interactions at the interface.
\\
In conclusion, we report the observation of exchange bias in the double perovskite
Sr$_2$FeCoO$_6$ which is a spin glass.
The claim of intrinsic exchange bias is supported through field-cooled hysteresis
measurements, dependence of exchange bias fields on cooling fields and temperature,
{\it training effect} etc.
Interface magnetic interactions between FM regions and SG domains
are believed to be the origin of observed exchange bias.
The underlying spin glass phase is further characterized through the conformation
with $AT$-line, {\it memory} and {\it aging} effects and magnetic relaxation.
\\
The authors acknowledge the Department of Science and Technology (DST),
India for the financial support for providing the facilities used in this
study (Grant No. SR/FST/PSII-002/2007) and (Grant No. SR/NM/NAT-02/2005).\\
* Authors contributed equally to this work.
%
\clearpage
\newpage
%
%
%
%
\clearpage
\newpage
%
%
%
\clearpage
\newpage
%
%
\clearpage
\newpage
%
%
%
%
\clearpage
\newpage
%
\begin{table}[!htb]
\caption{The parameters, $M_0$, $M_g$, $t_r$ and $n$ obtained by fitting the time-dependent magnetization at different wait times using eqn ~\ref{K}.}
\centering
\begin{tabular}{c c c c c} \hline
$t_w$ (s)             &  $M_0$ (emu/g)      &   $M_g$ (emu/g)     &  $t_r$ (s) & $n$\\ \hline\hline
300  & 0.009(3)   &  0.074(1)  & 1590(1)  & 0.43(1)   \\
1500  & 0.009(2)   &  0.073(2)  & 1671(2)  & 0.41(2)    \\
3000  & 0.0102(3)   &  0.071(2)  & 1705(1)  & 0.41(5)    \\ \hline\hline
\end{tabular}
\label{fit1}
\end{table}
\clearpage
\newpage

\begin{thebibliography}{10}


\bibitem{park_2011}
B.~G. Park, J.~Wunderlich, X.~Mart{\'\i}, V.~Hol{\`y}, Y.~Kurosaki, M.~Yamada,
  H.~Yamamoto, A.~Nishide, J.~Hayakawa, H.~Takahashi et al.,
\newblock {\em Nature Mater.}, \textbf{10}(5), 347 (2011).

\bibitem{nogues_1999}
J.~Nogu{\'e}s and  I.~K. Schuller.
\newblock {\em J. Magn. Magn. Mater}, \textbf{192}, 203 (1999).

\bibitem{liao_2010}
J.~-W. Liao, R.~K. Dumas, H.~-C. Hou, Y.~-C. Huang, W.~-C. Tsai, L.~-W. Wang, D.~-S.
  Wang, M.~-S. Lin, Y.~-C. Wu, R.~-Z. Chen, et al.,
\newblock {\em Phys. Rev. B}, \textbf{82}(1), 014423 (2010).

\bibitem{roshchin_2010}
I.~V. Roshchin, O.~Petracic, R.~Morales, Z.~P. Li, X.~Batlle, I.~K. Schuller.
\newblock US Patent 7,764,454 (2010).

\bibitem{stiles_2001}
M.~D. Stiles and R.~D. McMichael.
\newblock {\em Phys. Rev. B}, \textbf{63}, 064405 (2001).

\bibitem{skumryev_2003}
V.~Skumryev, S.~Stoyanov, Y.~Zhang, G.~Hadjipanayis, D.~Givord, and
  J.~Nogu{\'e}s.
\newblock {\em Nature}, \textbf{423}(6942), 850 (2003).

\bibitem{karmakar_prb_77_144409_2008}
S.~Karmakar, S.~Taran, E.~Bose, B.~K. Chaudhuri, C.~P. Sun, C.~L. Huang, and
  H.~D. Yang.
\newblock {\em Phys. Rev. B}, \textbf{77}, 144409, (2008).

\bibitem{tang_prb_73_174419_2006}
Y.~{-k} Tang, Y.~Sun, and Z.~{-h} Cheng.
\newblock {\em Phys. Rev. B}, \textbf{73}, 174419, (2006).

\bibitem{chen_prb_72_054436_2005}
X.~H. Chen, K.~Q. Wang, P.~H. Hor, Y.~Y. Xue, and C.~W. Chu.
\newblock {\em Phys. Rev. B}, \textbf{72}, 054436, (2005).

\bibitem{nogues_prl_76_4624_1996}
J.~Nogu\'{e}s, D.~Lederman, T.~J. Moran, and I.~K. Schuller.
\newblock {\em Phys. Rev. Lett.}, \textbf{76}, 4624 (1996).

\bibitem{ali_nature_6_70_2006}
M.~Ali, P.~Adie, C.~H. Marrows, D.~Greig, B.~J. Hickey, and R.~L. Stamps.
\newblock {\em Nature Mater.}, \textbf{6}, 70 (2007).

\bibitem{mukherjee_1996}
S.~Mukherjee, R.~Ranganathan, P.~S. Anilkumar, and P.~A. Joy.
\newblock {\em Phys. Rev. B}, \textbf{54}(13), 9267 (1996).

\bibitem{pradheesh}
R.~Pradheesh, Harikrishnan~S. Nair, C.~M.~N. Kumar, Jagat Lamsal, R.~Nirmala,
  P.~N. Santhosh, W.~B. Yelon, S.~K. Malik, V.~Sankaranarayanan, and
  K.~Sethupathi.
\newblock {\em J. Appl. Phys.}, \textbf{111}(5), 053905 (2012).

\bibitem{andreev_jac_260_196_1997}
S.~V. Andreev, M.~I. Bartashevich, V.~I. Pushkarsky, V.~N. Maltsev, L.~A.
  Pamyatnykh, E.~N. Tarasov, N.~V. Kudrevatykh, and T.~Goto.
\newblock {\em J. Alloys and Comp.}, \textbf{260}(1), 196--200 (1997).

\bibitem{patra_jpcm_21_078002_2009}
M.~Patra, M.~Thakur, K.~De, S.~Majumdar, and S.~Giri.
\newblock {\em J. Phys.: Condens. Matter}, \textbf{21}, 078002 (2009).

\bibitem{luo_2007}
W.~Luo and F.~Wang.
\newblock {\em Appl. Phys. Lett.}, \textbf{90}, 162515 (2007).

\bibitem{giri_jpcm_23_073201_2011}
S.~Giri, M.~Patra, and S.~Majumdar.
\newblock {\em J. Phys.: Condens. Matter}, \textbf{23}, 073201 (2011).

\bibitem{almeida_1987}
J.~R.~L. de~Almeida and R.~Bruinsma.
\newblock {\em Phys. Rev. B}, \textbf{35}, 7267 (1987).

\bibitem{shoemaker_prb_80_144422_2009}
D.~P. Shoemaker, E.~E. Rodriguez, R.~Seshadri, I.~S. Abumohor, and T.~Proffen.
\newblock {\em Phys. Rev. B}, \textbf{80}, 144422 (2009).

\bibitem{martinez_1998}
B.~Martinez, X.~Obradors, Ll.~Balcells, A.~Rouanet, and C.~Monty.
\newblock {\em Phys. Rev. Lett.}, \textbf{80}(1), 181 (1998).

\bibitem{palmer_1984}
R.~G. Palmer, D.~L. Stein, E.~Abrahams, and P.~W. Anderson.
\newblock {\em Phys. Rev. Lett.}, \textbf{53}(10), 958 (1984).

\bibitem{bohmer_1993}
R.~B\"{o}hmer, K.~L. Ngai, C.~A. Angell, and D.~J. Plazek.
\newblock {\em J. Chem. Phys.}, \textbf{99}(5), 4201 (1993).

\bibitem{chamberlin_1984}
R.~V. Chamberlin.
\newblock {\em Phys. Rev. B}, \textbf{30}(9), 5393 (1984).
\end{thebibliography}

\end{document}